\documentstyle[multicol,aps,graphics]{revtex}
\begin{document}
%\twocolumn[\hsize\textwidth\columnwidth\hsize\csname@twocolumnfalse\endcsname
\title{Multi-Choice Minority Game}
\author{Liat Ein-Dor$^1$, Richard Metzler$^2$, Ido Kanter$^1$ 
and Wolfgang Kinzel$^2$}
\address{
     $^1$ Department of Physics, Bar-Ilan University\\ 
       \mbox{~} Ramat Gan, 52900 Israel \\       
     $^2$ Institut f\"{u}r Theoretische Physik, Universit\"{a}t W\"{u}rzburg\\
      \mbox{~} Am Hubland, D-97074 W\"{u}rzburg, Germany
}
\maketitle

\begin{abstract}
The generalization of the problem of adaptive competition, known as 
the minority game, to the case of $K$ possible choices for each player
is addressed, and applied to a system of interacting perceptrons with input 
and output units of the type of $K$-states Potts-spins. 
An optimal solution of this minority game as well as the dynamic 
evolution of the adaptive strategies of the players are solved analytically 
for a general $K$ and compared with numerical simulations.
\end{abstract}
\begin{multicols}{2}

\section{Introduction}

Considerable progress in the theoretical understanding of market phenomena has
 been achieved by the study of the minority game. This prototypical model 
describes a system of agents interacting through a market mechanism 
\cite{Min2,Min3,Min4,Min5,Min6,Min7}. The game is based on the idea that the 
behavior of the agents is determined by the economic rule of supply and
demand. According to this rule, given the available options 
(such as buy/sell), an agent wins if he chooses the minority action. 
The research of this game has been focused on cases in which each agent
can choose between two options using its most efficient strategy,
 where the strategies remain unchanged throughout the game 
\cite{Min2,Min3,Min4,Min5,Min6,Min7}.
 However, 
in  the real world, many situations of interest involve more than two 
decision options as well as agents with dynamic strategies. Making decisions
like where to spend the summer vacation or which server to choose 
while surfing the web (or more generally, how to distribute
data traffic in computer networks \cite{Web}) are only two
 among many common problems with more than two options.
Therefore, it is tempting to investigate cases with more than two possible 
choices provided to agents with dynamic strategies.
 In a recent study of an extension in which each agent is
equipped with a neural network for making his decision \cite{Min1} 
 it was shown that a certain updating rule
of the strategies of the agents improves the efficiency of the market, 
which is measured by the global profit of the agents. In this paper we
generalize the aforementioned work to a multi-choice minority game, namely
 a game with general $K$ decision states.

The multi-choice minority game consists of $N$ players (agents)
and $K$ possible decisions. In each step, each one of the players 
chooses one of the $K$ states, aiming to choose the state with 
the smallest number of agents. For example, a situation may arise, 
in which there are several possible roads which lead from place A
to place B, and each  driver who wants to get from A to B chooses one
 of the available roads.
 Because drivers want to avoid traffic jams, they try to choose the least
 traveled roads,
assuming that all the roads are of the same length.
 Similarly, one usually
prefers to go to the bar with the smallest number of people in it.  
 Occurring over and over again, the minority decisions in these and other 
similar situations generate time series
whose term at time $t$, $x_t$, has an integer value between $1$ and $K$
according to the minority decision. In the original game, the information
 provided to each player is the history vector of size $M$, whose components
 are the last $M$ minority states.
 
The paper is organized as follows: In section \ref{SecModel}
a multi-layer neural network and the dynamic evolution  of its weights are
introduced. 
 For the clarity 
of the rest of the paper which is somewhat technical we briefly discuss the 
main findings and results. 
In section \ref{SecRandom} the reference case of players with
random strategies is solved                  
analytically. 
In section \ref{SecOptimal} the global profit of the players 
for the network with 
optimal strategies (weights) is solved analytically in the thermodynamic
limit, and shown to be superior to a random decision. The analytical 
results are compared with  simulations on finite systems.
In section \ref{SecDynamics},
the suggested updating rules for the weights are examined analytically and 
are found to saturate asymptotically the optimal global profit.
Finally, section \ref{SecSummary} is devoted to a short 
summary and an outlook. 

\section{ The model }
\label{SecModel}
While many strategies for the multi-choices minority game 
are conceivable, we study the following model which 
uses neural networks: 
% or something along these lines
each one of the $N$ players is represented by a perceptron 
of a size $M$. The weights belonging to the $i$th player are $\{w_{ij}\}$ where $j=1, . . ., M$. 
All $N$ perceptrons have a common input which consists of $M$ components 
$x_1,~...,~x_M$, where each one of the components can take one of the 
$K$ integers, $1,~2,~...,~K$, with equal probability.

The dynamics are defined by the following steps.
In the first step, each one of the perceptrons calculates the $K$ induced 
local fields.
For instance, the  field $h_{im}$ induced by the $m$th state on player $i$
is defined as the summation over all weights belonging to
the $i$th perceptron with input equal to $m$:
\begin{equation}
h_{im}=\sum_{j=1}^Mw_{ij}\delta_{x_j,m}.
\end{equation}
In the second step, each player chooses its state, $\{\sigma_i\}$, 
following the maximal induced field: 
\begin{equation}
\sigma_i=\{k_1\mid\max_{m=1,..,K}h_{im}=h_{ik_1}\}.
\end{equation}
\noindent where $\sigma_i$ is 
the output (chosen state) of $i$th perceptron.
In the third step, the occupancy of each state is calculated:
\begin{equation}
N_\rho=\sum_{j=1}^N \delta_{\sigma_j,\rho},
\end{equation} 
\noindent where it is clear that $\sum_{\rho} N_\rho =N$.
The output $min$ of the network is the minority decision 
\begin{equation}
min=\{\rho \mid\min_{m=1,..,K}N_m = N_\rho\},
\end{equation} 

The game can also be represented by a feedforward network $M:N:1$ 
($M$ input units, $N$ hidden units and $1$ output). 
All units (input, hidden, output)
 are represented by $K$-states Potts-spins. The weights $\{w_{ij}\}$ are from 
the input units to the hidden units, and the weights from the hidden units to 
the output are all equal to $-1$. The dynamics of  hidden
and output units are similar to zero temperature dynamics of Potts-spin
systems \cite{Potts1,Potts2}, following the maximal induced field.
The free parameters in our game are the $MN$ weights,
$\{w_{ij}\}$, from the input to the hidden units. Their values will be
determined by the strategy adopted by each one of the players. Our local 
dynamic rules are based on the generalization of the on-line 
Hebbian learning rule for $K=2$ \cite{Min1} to general $K$-states 
Potts model with the following  updating rule;
\begin{equation}
w_{ij}^+=w_{ij}+{{\eta}\over{M}}(K\delta_{x_j,\min}-1)
\label{update}
\end{equation}
where $\eta$ is the learning rate  
and the sign $+$ indicates the next time step. 
Note that all agents use the same rule for updating their
strategy. 

The score of the game is determined similarly to the Ising case.
 Players belonging to the minority ($N_{min}$ players) 
gain $Q_+$, while the other $N-N_{min}$ players 
gain $Q_-$, where $Q_+ > Q_-$. 
Note that in most previous works $Q_+$ was chosen 
to be $1$ and $Q_-$ was chosen to be either
$0$ or $-1$. The global profit in such cases is
\begin{equation}
U=Q_-N+(Q_+-Q_-)N_{min}.
\label{profit}
\end{equation} 
\noindent It is clear that the maximization of the global profit $U$ is
equivalent
to the maximization of $N_{min}$, which is bounded from above by $N/K$.
Note that in the Ising case each player belongs either to the minority or to 
the 
majority, where in the Potts case the situation is more complex.
The score may depend on the exact values of $\{N_i\}$ 
(the score decreases with $N_\rho$),
hence the total profit  $U=U(\{N_i\})$. 
In such a case the  maximization of the total profit may differ from 
the maximization of $N_{min}$,  and will be discussed briefly in the end of 
this paper.

Before we turn to discuss  the guideline of the 
derivation of the results, which are more involved than for the Ising case, 
let us present the main results:
(a) The score and the dynamics are formulated analytically for general
$K$, the number of possible decisions. Exact results are obtained for
$K \le 6$ and asymptotically for $K \rightarrow \infty $. 
Results for intermediate
 values of $K$ are obtained from simulations.
(b) A relaxation to the optimal score is achieved for small learning rates.
(c) Regarding the optimal case, the deviation of minority group size from 
$N/K$  is found to be non-monotonic with $K$.   
(d) The total score is independent of the size of the history ($M$, the size
of the input) available for the agents. 
(e) All agents are using the same type of dynamic strategy 
and gain on average 
(over time) the same profit.  
Our system does not undergo a phase transition to a state where the symmetry
among the agents is broken into losers and winners \cite{Min5,Min6}.
Throughout the investigation of the game we assume that the memory size
$M$ is larger than the number of players $N$ (otherwise the 
completely symmetric Potts configuration is geometrically impossible).
Albeit, simulations of the same dynamic 
for systems with $M<N$ show even better results for the global profit.

\section{ The random case }
\label{SecRandom}
In case where 
the maximization of the global profit $U$ is identical
to the maximization of $N_{min}$, the quantity of interest is 
\begin{equation}
<{\epsilon_{min}}^2>= {1\over{N}}< ( N_{min} - N/K )^2 >,
\end{equation} 
\noindent where the symbol $<~~ >$ indicates an average over input 
patterns, and 
$N/K$ is the average number of players in each state. Note that in our 
calculations the input vector presented to the players 
at each step of the game consists of random components\cite{Min5,Min1},
instead of the true history.
Nevertheless, simulations indicate that the system behavior is only
slightly affected by the randomness of the inputs and the game properties 
remain similar.

\begin{figure}
 \begin{center}
   \resizebox{0.95 \columnwidth}{!}
   {\rotatebox{270}{\includegraphics{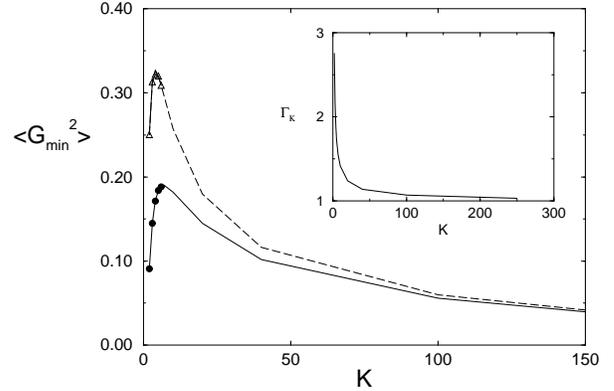}}} 
 \end{center}
                    
\caption{
Simulations for $<{\epsilon_{min}}^2>$ as a function of $K$ for both the 
optimal case $R=-{1\over{N-1}}$ (solid curve) and the random case $R=0$ (long 
dashed curve). Analytical results up to $K=6$  
agree with simulations for both $R=-{1\over{N-1}}$ (filled circles) and 
$R=0$ (triangles). Inset: 
$\Gamma_K=<{\epsilon_{min}}^2>_{R=0}/<{\epsilon_{min}}^2>_{R=-{1\over{N-1}}}$ 
as 
a function of $K$.  
}
\label{fig1}
\end{figure}

For random players, each weight (among the $MN$ weights $\{w_{ij}\}$) 
is chosen from a given unbiased distribution and a variance $1/M$.
Hence, the distribution of the overlap $R$ between   
weights belonging to any two
players $\rho$ and $\phi$
\begin{equation}
R_{\rho \phi}= \sum_{j=1}^M w_{\rho j}w_{\phi j}
\end{equation}
\noindent is a Gaussian with zero mean and 
variance $1/M$.
In the thermodynamic limit and for $M>N$, one can show that in the  
leading order the overlap between each pair is an 
independent random variable. For random players and $K=2$ one finds 
$<\epsilon^2> = <\sum_{\rho}(N_{\rho} -N/K)^2/(NK)> = 1/4$; 
 however, for general $K$ even the derivation of a similar quantity is 
 non-trivial. The two cornerstones of the calculations below are the
probability of a microscopic configuration $P(\{\sigma_i\})$, and
the degeneracy $D(\{N_{\rho}\})$ of a macroscopic configuration $\{N_{\rho}\}$,
which is given by the multinomial coefficient:
\begin{equation}
  D(\{N_{\rho}\}) = \frac{N!}{\prod_{\rho} N_{\rho}!}. \label{Deg1}
\end{equation}
In the large $N$ limit, the typical deviation of the size of each
 group from $N/K$ is expected to scale with $\sqrt{N}$. Hence we define:
\begin{equation}
N_\rho =N/K + \epsilon_\rho \sqrt{N}
\end{equation}
\noindent where it is clear that $\sum_\rho \epsilon_\rho =0$ and
without the loss of generality we assume 
$N_{min}=N_1 \le N_\rho~~\forall\rho >1$.
Applying the Stirling approximation to Eq. (\ref{Deg1})
yields the degeneracy as a function of $\{ \epsilon_\rho \}$:
\begin{equation}
D_K(\{\epsilon_\rho\})\sim K^N \exp({-{K \over 2}\sum_{\rho=1}^{K}
 \epsilon_\rho^2 })
\delta ( \sum_{\rho=1}^K \epsilon_\rho ). \label{Deg2}
\end{equation}
If the average over $R_{\rho\phi}$, which we denote by $R$, is $0$, 
the agents make their choice independently and randomly, so
each microscopic configuration has the same probability $P_K = (1/K)^N$.
Now the average over $\epsilon_{min}^2$ can be evaluated:
\begin{equation}
 <\epsilon_{min}^2>_{R}={
\int_{-\infty}^{0}
{\epsilon_{1}}^2d\epsilon_{1} \Pi_{\rho >1} 
\int_{\epsilon_{1}}^{\infty}
d\epsilon_{\rho} D_K(\{\epsilon_{\rho}\})P_K(\{\epsilon_{\rho}\})
\over
\int_{-\infty}^{0}
d\epsilon_{1} \Pi_{\rho >1} 
\int_{\epsilon_{1}}^{\infty}
d\epsilon_{\rho} D_K(\{\epsilon_{\rho}\})P_K(\{\epsilon_{\rho}\}) }.  
\label{epsmin}
\end{equation}
The quantity $<{\epsilon_{min}}^2>_{R=0}$ was calculated 
numerically for $K=3,~4,~5,~
6$ and found to be equal to  
$\sim 0.313,~0.322,~0.320,~0.309$, respectively (see Fig. \ref{fig1}). 
Results obtained from simulations with $N=5000$ and $K\le 6$ are in an
 excellent agreement with Eq. (\ref{epsmin}).
For $K > 6$ the reported results in Fig. \ref{fig1} were derived 
only from simulations and are in an excellent agreement with the asymptotic 
behavior of Eq. (\ref{epsmin}), $<\epsilon_{min}^2>_{R=0}\sim 2\log(K)/K$.
Another quantity of interest is the average deviation 
of the average number of players in each state from $N/K$, 
$<\epsilon^2> =<{1\over{K}}\sum_\rho \epsilon_\rho^2>$. Similarly to 
Eq. (\ref{epsmin}), this quantity can be derived analytically  
and gives 
\begin{equation}
<\epsilon^2>_{R=0}={{K-1}\over{K^2}}.
\label{epssqr}
\end{equation}

\section{ The optimal case }
\label{SecOptimal}
So far we have compared $<{\epsilon_{min}}^2>$ and $<{\epsilon}^2>$ 
for random players, 
where the average overlap is zero. 
Without breaking the symmetry among the players, 
the weights can be represented by 
$N$ weight vectors which are symmetrically spread around
their center of mass. More precisely, we denote the weight 
vector of the $i$th perceptron as ${\bf w}_i$, and assume that 
it can be expressed as 
\begin{equation}
{\bf w}_i=  {\bf C} +{\bf g}_i, 
\end{equation}
\noindent where the center of mass ${\bf C} \equiv {1 \over N} \sum_{i}
 {\bf w}_i$, and $\{ {\bf g}_i \}$ are $N$ unit vectors of rank $M$ obeying
 the symmetry
\begin{equation}
{\bf g}_i \cdot {\bf g}_j= (1+{{1}\over{N-1}})\delta_{ij} + {{1}\over{N-1}}. 
\end{equation}
\noindent Hence, the total profit and $N_{min}$ are functions of only 
one parameter, $C$.
It is clear that the maximization of the total profit or $N_{min}$ 
(as for the case $K=2$) is
obtained when $C=0$, which is the maximal achievable
homogeneous repulsion  among $N$ vectors of rank $M > N$.
The repulsion is the natural tendency of each player in the minority game,
since the goal is to act differently than other players. Without 
a cooperation which breaks the players into sub-groups, the maximal 
 homogeneous repulsion is $R=-1/(N-1)$.

The two questions of interest are the following:
(a) What is $<\epsilon^2>$ and $<{\epsilon_{min}}^2>$ as a function of $K$ 
for the optimal solution, $C=0$ 
and $R=-1/(N-1)$? (b) Is the optimal solution achievable by
local dynamic rules for each one of the players?
We first examine the former question regarding the optimal solution, and 
then we turn to study the dynamic behavior of the players.

The average deviation of the number of players in each state from $N/K$
at $C=0$ and for $R=O(1/N)$ can be
calculated analytically. The main idea is that 
this quantity can be calculated similarly to Eq. (\ref{epsmin}),
 or via
$<{\epsilon}^2>={1/(NK)}<(\sum_{\rho=1}^K\sum_{i=1}^N\delta_{\sigma_i,\rho}
-{N/K})^2>$. 
The simplification of 
the later expression is such that an average over  
only a pair of players has to be done. The result as a function of $K$ gives
\begin{equation}
<{\epsilon}^2>_{R}={{K-1}\over{K^2}}+R(N-1)(K-1)K\mu ,
\label{epssqrR}
\end{equation}
\noindent where $\mu=[\int_{-\infty}^{\infty}
{e^{-h^2}\over{2\pi}}{(1-H(h))}^{K-2}dh]^2$ and  
$H(x)=0.5\;\mbox{erfc}({x\over\sqrt{2}})$. 

Regarding the optimal score, the quantity of a particular interest 
is $<{\epsilon}_{min}^2>_{R={{-1}\over{N-1}}}$. 
This quantity has to be compared with  $<{\epsilon}_{min}^2>_{R=0}$ 
in order to estimate the improvement in the average global 
gain relative to the random case. Note that the calculation of 
Eq. (\ref{epsmin})
for  $R\neq 0$ is nontrivial since 
$P_K(\{\epsilon_{\rho}\})$ is no longer independent of 
the configuration $\{\epsilon_{\rho}\}$.
However, we can overcome this difficulty in the following way.
For $R=O(1/N)$ one can show that in the leading order 
$P_K(\{\epsilon_\rho\})$ has the same form as $D_K(\{\epsilon_\rho\})$:
\begin{equation}
P_K(\{\epsilon_\rho\}) \sim ({1/K})^N \exp({-{A(R)}\sum_{\rho=1}^{K} 
\epsilon_\rho^2 })\delta ( \sum_{\rho=1}^K \epsilon_\rho ),
\label{PKeps}
\end{equation}
where the exact value of $A(R)$ is unknown.
The observation that both 
$P_K(\{\epsilon_\rho\})$ and $D_K(\{\epsilon_\rho\})$ have 
the same dependence on $\{\epsilon_\rho\}$ 
(Eqs. (\ref{Deg2}) and (\ref{PKeps})) 
indicates that 
the  ratio $<\epsilon_{min}^2>/<\epsilon^2>$ is independent of $R$
if $R=O(1/N)$, 
and in particular:
\begin{equation}
{<{\epsilon}^2>_{R=0}\over{<{\epsilon_{min}}^2>}_{R=0}}=
{<{\epsilon}^2>_{R=-{1\over{N-1}}}\over{<{\epsilon_{min}}^2>}_
{R=-{1\over{N-1}}}}
=\beta_K.  
\label{ratio}
\end{equation}
This  property  can be easily derived by rescaling $\epsilon_{\rho}\to 
\sqrt{A(R)}\epsilon_{\rho}$ in the integral representation 
(Eq. (\ref{epsmin})) of
each one of the four terms in Eq. (\ref{ratio}).
The same prefactor appears both in the denominator 
and in the numerator, and
the dependence of $\beta_K$ on $R$ via $A(R)$ is cancelled out.
Using Eq. (\ref{ratio}), 
${<{\epsilon_{min}}^2>}_{R=-{1\over{N-1}}}$ can be obtained 
indirectly from the knowledge of the other three terms,
which
are given by Eqs. (\ref{epsmin}), (\ref{epssqr}), and (\ref{epssqrR}).
 Results for
${<{\epsilon_{min}}^2>}_{R=-{1\over{N-1}}}$ are presented in Fig. \ref{fig1}. 
In order to confirm our analytical results 
we performed simulations for the optimal case, Eqs. (\ref{epssqrR}) 
and (\ref{ratio}),
 with $C=0$. The simulations were done in two stages. 
In the first stage, $N$ normalized vectors of rank $M$, obeying the 
constraints that the overlap among each pair is equal to $-1/(N-1)$, are 
generated using a recursive process. The details of the algorithm will be 
given elsewhere
 \cite{Min8}.
In the second stage, 
$<{\epsilon_{min}}^2>$ and $<{\epsilon}^2>$ were averaged over about $10^5$ 
randomly chosen inputs for a system with $N=400$ and $M=5000$.
An excellent agreement between simulations and analytical 
results was obtained (see Fig. \ref{fig1}). The improvement in the global gain 
can be measured by the ratio $\Gamma_K=<{\epsilon_{min}}^2>_{R=0}/
<{\epsilon_{min}}^2>_{R={-1\over{N-1}}}$. This ratio
decreases monotonically with $K$ such that its maximal value 
$\Gamma_2 = 2.7548$ and for $K\to \infty$ $\Gamma_K \to 1$ 
(inset of Fig. \ref{fig1}). 

\section{ The dynamics which lead to the optimal solution }
\label{SecDynamics} 

So far we derived the properties of the optimal solution  for 
different values of $K$. Now we are turning to the second question: is 
the optimal solution achievable by local dynamic rules 
(Eq. (\ref{update}))?
After averaging Eq. (\ref{update}) over $j$ and in the limit 
where the number of examples, $\alpha{M}$,  scales with 
the number of input units $M$,
one can find the following equation of motion for the center of mass
\begin{equation}
{{dC^2}\over{d\alpha}}=2\eta{K}<\sum_jC_j\delta_{x_j,min}>+\eta^2(K-1),
\label{Cfix}
\end{equation}
where $<~~ >$ denotes an average over the random examples. 
\noindent For large $M$, in the leading order each input 
vector divides each weight  vector into $K$ equal groups of size $M/K$.
The minority state is the one whose group of weights
gives the minimal sum.
Using Eq. (\ref{Cfix})
and $M,N\to\infty$, 
$<\sum_jC_j\delta_{x_j,min}>$ is the average minimal sum of a set 
of $M/K$ center of mass components,  $\{C_j\}$. 
These $M/K$ quantities are random variables 
with zero mean and variance $C^2/M$ ( $<\sum_{j=1}^{M/K}C_j>=0$ and
$<(\sum_{j=1}^{M/K}C_j)^2>=C^2/K$). 
One can find
that $<\sum_j{C_j\delta_{x_j,min}}>$ is equal to 
\begin{equation}
{C\over{2}}(K-1)\sqrt{K\over{\pi}}\int_{-\infty}^{\infty}
{e^{-y^2/2}\over\sqrt{2\pi}}[H({y\over{\sqrt{2}}})]^{K-2}dy.
\label{Cavg}
\end{equation} 

\begin{figure}
 \begin{center}
   \resizebox{0.95 \columnwidth}{!}
    {\rotatebox{270}{\includegraphics{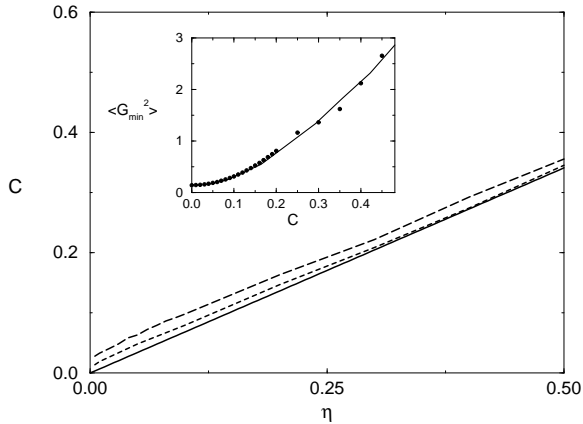}}} 
 \end{center}
\caption{
C as a function of $\eta$ for $K=3$. Analytical results (solid line) and 
simulations for  $N=103$, $M=200$ (long-dashed), 
and for $N=400$, $M=403$ (dashed-line).
Inset: $<{\epsilon_{min}}^2>$ as a function of $C$ for $K=3$. Analytical 
results (solid line) and simulations for $N=103$, $M=200$ (filled circles).
}
\label{fig2}
\end{figure}

\noindent Hence, for a given $K$, Eqs. (\ref{Cfix}) and (\ref{Cavg})
 indicate a linear relation 
between the fixed point value of $C$ and the learning rate $\eta$
 with corrections of $O(1/\sqrt{N})$. 
As $\eta \to 0$, $C \to 0$ and the system approaches the optimal 
configuration.  
The interplay between $C$ and $\eta$ was
 confirmed by simulations, where finite size effects decay as the size of 
the system becomes larger. 
This effect is depicted in Fig. \ref{fig2} for $K=3$.
The explicit dependence of $<{\epsilon_{min}}^2/N>_{R}$ on $C$ can be 
found for $R\sim O(1/N)$ via the relation
\begin{equation}
R={{C^2-
{1\over{N-1}}}\over{C^2+1}}.
\end{equation} 
\noindent
Results of simulations for $<{\epsilon_{min}}^2>_{R}$ as a function of $C$ 
for $N=103$ and $M=200$ are presented in the inset of Fig. \ref{fig2}. 
An excellent 
agreement between the analytical prediction and simulations was obtained in 
the regime of $C\sim O(1/\sqrt{N})$ (corresponding to $R\sim O(1/N)$). 

Note that although the global gain $U$ which corresponds to the Boolean
case is
 monotonic with $K$, the non-monotonic behavior of ${\epsilon_{min}}$ implies
 that for non-Boolean cases non-monotonic behavior of $U$ may be obtained.

\section{ summary and outlook }
\label{SecSummary}
In this paper we introduced a generalization of the minority game to
the case of multi-choice. The problem was applied to a multilayer network
 with updating rules for the weights (strategies). Static and dynamic
 properties of the strategies were solved analytically for various $K$'s
 and were found to be in a good agreement with simulations on finite systems.
This modification of the minority game to the case of multi-choice 
open a manifold of new questions, which certainly deserve future research.
 We have chosen two of those questions to briefly discuss here.

Firstly, as we have pointed out before, the function according to which 
the profit is 
awarded is not necessarily Boolean as in Eq. (\ref{profit}). 
In fact, the model is more 
realistic when the profit of a player is related to the size of his group, as 
well as to the size of the other groups \cite{Min9}. 
Our analysis can be applied to these cases if the 
maximization of the global gain is equivalent to the maximization of the
 minority group. 
However, other scores may not fulfill this required condition. In these cases,
 it has to be determined whether the optimal symmetric configuration 
remains the maximal repulsion.

Secondly, the other strategies for the minority game that have been studied 
can be generalized to multi-choice situations in a straightforward manner:
in the original game \cite{Min2,Min3,Min6,Min7} where each player has 
several decision tables, each table entry is now a value between $1$
and $K$. In Johnson's stochastic strategy \cite{Johnson1,Johnson2},
each player has 
a probability of choosing the outcome that was successful the last time, or to
pick one of the others with equal probability. In the strategy of Reents 
\cite{Reents},
players who were not in the minority could switch to some other action with 
a small probability in the next time step. Similarly, other conceivable
strategies can also be generalized. Preliminary checks 
imply that all these modified strategies show similar behavior compared
 to that of the 
binary-choice game, even though their theoretical treatment probably 
becomes more involved. While outcomes of these games certainly have to be 
measured against the reference values given in Eqs. (12) and (13), it is not
clear under what circumstances relations like Eq. (18) hold for other 
strategies.  

I. K., W. K. and R. M. acknowledge a partial support by GIF.

\end{multicols}{2}

\clearpage

\end{document}